# Displaced rotations of coherent states

Sergey A. Podoshvedov

*Department of General and Theoretical Physics, South Ural State University, Lenin Av. 76, Chelyabinsk, Russia*

We propose an approach that enables to use it for construction of rotations of coherent states, in particular, it gives a possibility to construct Hadamard gate for the coherent states. Our approach is based on representation of arbitrary one-mode pure state in free-travelling fields, in particular superposition of coherent states (SCSs), in terms of displaced number states with arbitrary amplitude of displacement. Studied optical scheme is based on alternation of photon additions and displacement operators (in general case, $N$ – photon additions and $N-1$ – displacements are required) with seed coherent state to generate both even and odd displaced squeezed SCSs (DSSCSs) regardless of number of used photon additions. It is shown the studied optical scheme is sensitive to seed coherent state provided that other parameters are invariable. Output states approximate with high fidelity either even squeezed SCS or odd SCS shifted relative each other by some value. It enables to construct local rotation operator for coherent states, in particular, Hadamard gate being mainframe element for quantum computation with coherent states. The effects deteriorating quality of output states are considered.

## I. INTRODUCTION

The development of the quantum theory of light has deepened our understanding of nonclassical properties of optical fields. But realization in the laboratory of schemes for the generation of specific nonclassical quantum states is one of the most exciting challenges to the researches. Recently, superpositions of coherent states in free-traveling fields (SCSs) [1, 2] have attracted special attention due to their remarkable usefulness. The superposition of two coherent states (i.e., most classical) with opposite phases [3] exhibits both some properties similar those of statistical mixtures and typical interference features. So, one of the quadrature-component distributions of SCSs shows two peaks that change their mutual distance in dependency on amplitude of coherent fields, whereas an oscillatory behavior is observed in another quadrature-component distribution [3]. Note that such a behavior takes place only for large amplitudes of coherent states composing SCSs when somehow macroscopically distinguishable outcomes are observed by a homodyne measurement. Note also negative values in the Wigner functions of the SCSs [2] being also manifestation of their nonclassical properties. The SCSs enable one to perform many interesting studies for fundamental tests of quantum theory [4-6]. Furthermore, it has been found that SCSs are useful for various applications in quantum information processing [7-12].

In spite of the manifold usefulness of the SCSs, until recently, there was not serious progress in the generation of the SCSs. There have been schemes to generate such SCSs using strong nonlinearities [1,13] or photon number resolving detectors [14] which are hardly feasible with current level of technology. Recently, more realistic schemes have been proposed by different authors [15-17]. So, a simple observation that odd SCS with small amplitude $\leq 1.2$ is well approximated by squeezed single photon was made in [15]. Also it was noted that squeezed single photon can be obtained by subtracting (or adding) one photon from pure squeezed vacuum [18]. Theoretical analysis of added/subtracted squeezed vacuum states has been performed in [19]. Single-photon-subtracted squeezed states, which are close to SCSs, have been generated in [20]. Squeezed SCS with size of the states $\approx 1.6$ was generated and detected [21]. It may be suited for fundamental tests and quantum information processing despite their squeezing [22]. Subsequent steps were aimed at study of two-photon added/subtracted squeezed vacuum states [23,24]. A scheme using time separated two-photon subtraction to generate SCS of large amplitude was experimentally demonstrated in [25].

In this paper, we are interested in finding general approach to analysis of such problems to use it for both generation of squeezed SCSs with larger amplitude and higher fidelity and construction of local rotations with coherent states that may open new possibilities for the quantum computation with coherent states. To develop the general approach, we make use of possibility to decompose arbitrary one-mode pure state in free-travelling fields in terms of displaced number states with arbitrary amplitude of displacement [26]. The decomposition is possible since the set of the displaced states is complete and they are orthogonal with respect to the inner product. The displaced number states were proposed to use for the dense coding [27] and quantum key distribution [28]. Note the displaced vacuum is a coherent state while a displaced single photon has been experimentally realized in [29]. It is interesting to note it is possible to express microscopic states, for example vacuum, as superposition of macroscopic states that is not subject of the investigation. Here, we consider optical scheme with alternate single photon additions and displacements [30] and show it can generate any type (even, odd) of squeezed SCSs, regardless of the number of added photons, shifted relative each other by some quantity. It enables to adjust the optical scheme for construction of local rotation and, in particular, Hadamard gate with coherent states. Corresponding optimal parameters of the scheme are collected in tables.

## II. SCSs IN TERMS OF THE DISPLACED NUMBER STATES

We define regular rotated SCSs as
$$|SCS_Q(\alpha_{SCS})\rangle = N_Q(\alpha_{SCS}) \left( \cos Q |0, \alpha_{SCS}\rangle + \sin Q |0, -\alpha_{SCS}\rangle \right), \quad (1)$$
where
$$N_Q(\alpha_{SCS}) = 1 / \sqrt{1 + 2\cos Q \sin Q \exp(-2|\alpha_{SCS}|^2)}$$
is a normalization factor and, in framework of our notations, $|0, \pm\alpha_{SCS}\rangle$ is a coherent state with amplitudes $\pm\alpha_{SCS}$, respectively. We call a state (1) regular $Q-$ SCS. If we take $Q = \pi/4$ or $Q = -\pi/4$, then we deal with regular even ($|SCS_+(\alpha_{SCS})\rangle$) and odd ($|SCS_-(\alpha_{SCS})\rangle$) SCSs, respectively. We accept $\alpha_{SCS} > 0$ throughout the paper. "Taking $\alpha_{SCS} > 0$ real" means that the field is in phase with the local oscillator which is used for qubit measurement and to make the displacements required for some of the gates. Displaced $n$ photon state, where $n$ is an arbitrary number, is defined as
$$|n, \alpha\rangle = \hat{D}(\alpha)|n\rangle,$$
where $D(\alpha) = \exp(\alpha a^+ - \alpha^* a)$ is a displacement operator, $a$ ($a^+$) is the bosonic annihilation (creation) operator, and $|n\rangle$ is a number state. In particular, $|0, \alpha_{SCS}\rangle = \hat{D}(\alpha_{SCS})|0\rangle$ (Eq. (1)) is a displaced vacuum state or the same coherent state of an amplitude $\alpha_{SCS}$.

Infinite set of the displaced number states $|l,\alpha\rangle$ ($l=0,1,2,...,\infty$), where $\alpha$ is an arbitrary number, is complete that enables to decompose any one-mode state on the basic states. We are going to call such a decomposition of arbitrary one-mode state $\alpha$ – representation of the state. Now, since we are interested in local transformations of the coherent states, we give the following $\alpha$ – representation of regular $Q$ – SCS that we are going to denote $|SCS_Q(\alpha_{SCS},\alpha)\rangle$ (Appendix A, formulas (A4) and (A5))

$$|SCS_Q(\alpha_{SCS},\alpha)\rangle = N_Q(\alpha_{SCS})$$
$$\exp(-(\alpha_{SCS}^2+|\alpha|^2)/2)D(\alpha)$$
$$\sum_{l=0}^{\infty}\frac{\alpha_{SCS}^l}{\sqrt{l!}}\begin{pmatrix}\cos Q\exp(\alpha_{SCS}\alpha^*)\\(1-\alpha/\alpha_{SCS})^l+\\\sin Q\exp(-\alpha_{SCS}\alpha^*)(-1)^l\\(1+\alpha/\alpha_{SCS})^l\end{pmatrix}|l\rangle \quad (2)$$

Two variables $\alpha_{SCS}$ and $\alpha$ are used for notation of arbitrary $\alpha$ – representation of the SCSs $|SCS_Q(\alpha_{SCS},\alpha)\rangle$ unlike direct definition of the SCSs $|SCS_Q(\alpha_{SCS})\rangle$ (Eq. (1)), where $\alpha_{SCS}$ means an amplitude of the SCS and $\alpha$ is an amplitude of the complete set of the displaced number states used for the decomposition of the SCS. Thus, it is worth distinguishing direct definition of the regular SCS (Eq. (1)) and its $\alpha$ – representation in framework of our notations. $\alpha$ – representation of the SCS is turned out to be very useful mathematical tool that is used for construction of the states. In particular, if we take $\alpha=0$, then we deal with number state representation (or the same $0$ – representation in framework of our notations) of the even/odd SCSs [3]

$$|SCS_+(\alpha_{SCS},0)\rangle = 2N_+(\alpha_{SCS})$$
$$\exp(-2|\alpha_{SCS}|^2)\sum_{n=0}^{\infty}\frac{\alpha_{SCS}^{2n}}{\sqrt{(2n)!}}|2n\rangle, \quad (3a)$$

$$|SCS_-(\alpha_{SCS},0)\rangle = 2N_-(\alpha_{SCS})$$
$$\exp(-2|\alpha_{SCS}|^2)\sum_{n=0}^{\infty}\frac{\alpha_{SCS}^{2n+1}}{\sqrt{(2n+1)!}}|2n+1\rangle. \quad (3b)$$

Since superpositions (3a), (3b) involve either even or odd number states, they are called even and odd SCSs, respectively. It is natural to use the same terms for the regular SCSs at general $\alpha$ – representation with $\alpha \neq 0$ (Eq. (2)) despite they involve both even and odd displaced number states by analogy with conventional terminology.

Consider optical scheme in Fig. 1(a) that can be used for the SCSs construction. The optical scheme starts with initial coherent state $|0,\alpha_{In}\rangle$ and consists of set of alternate photon additions and displacement operators [26,30]. In general case, the optical scheme in Fig. 1(a) involves $N$ photon additions and $N-1$ displacement operators. We are going to consider only partial cases with $N=1,2$ photon additions and $0,1$ displacement operators. An optical scheme with only one photon addition is a simplest as intermediate displacement operator is not used. We have in the case

$$a^+|0,\alpha_{In}\rangle = a^+D(\alpha_{In})|0\rangle =$$
$$D(\alpha_{In})D^+(\alpha_{In})a^+D(\alpha_{In})|0\rangle =$$
$$D(\alpha_{In})(a^+ + \alpha_{In}^*)|0\rangle =$$
$$\alpha_{In}^*\sqrt{1+1/|\alpha_{In}|^2}D(\alpha_{In}) \quad (4a)$$
$$(|0\rangle+|1\rangle/\alpha_{In}^*)/\sqrt{1+1/|\alpha_{In}|^2} =$$
$$\alpha_{In}^*\sqrt{1+1/|\alpha_{In}|^2}D(\alpha_{In})|\Psi_{\pm 1}\rangle$$

where

$$|\Psi_{\pm 1}\rangle = (|0\rangle+a_{\pm 1}|1\rangle)/\sqrt{1+|a_{\pm 1}|^2} =$$
$$(|0\rangle+|1\rangle/\alpha_{In}^*)/\sqrt{1+1/|\alpha_{In}|^2}, \quad (4b)$$

provided that $\alpha_{In} = 1/a_{\pm 1}^*$, is a simplest ($\pm 1$) "half-finished product" of the even/odd SCSs shifted by $\alpha_{In}$ and we made use of the following relation $D^+(\alpha)a^+D(\alpha)=a^++\alpha^*$ [31]. Consider the case of $N=2$, where "half-finished product" $|\Psi_{\pm 2}\rangle$ of the SCSs is given by

$$|\Psi_{\pm 2}\rangle = \frac{|0\rangle + a_{\pm 1}|1\rangle + a_{\pm 2}|2\rangle}{\sqrt{1+|a_{\pm 1}|^2+|a_{\pm 2}|^2}}. \quad (5a)$$

Displaced version of the states (5a) can be constructed using two photon additions with one intermediate displacement operator shifting by $\alpha_1$ (Fig. 1(a)) as

$$a^+D(\alpha_1)a^+|0,\alpha_{In}\rangle =$$
$$a^+D(\alpha_1)a^+D(\alpha_{In})|0\rangle =$$
$$\exp(i\phi)D(\alpha_1+\alpha_{In})D^+(\alpha_1+\alpha_{In})$$
$$a^+D(\alpha_1+\alpha_{In})D^+(\alpha_{In})a^+D(\alpha_{In})|0\rangle =, \quad (5b)$$
$$\exp(i\phi)D(\alpha_1+\alpha_{In})(a^++(\alpha_1+\alpha_{In})^*)$$
$$(a^++\alpha_{In}^*)|0\rangle = \exp(i\phi)D(\alpha_1+\alpha_{In})$$
$$\begin{pmatrix}\alpha_{In}^*(\alpha_1^*+\alpha_{In}^*)|0\rangle + (\alpha_1^*+2\alpha_{In}^*)|1\rangle \\ +\sqrt{2}|2\rangle\end{pmatrix}$$

where $\phi$ is some general phase shift and normalization factor is left out. Expression (5b) is a state $|\Psi_{\pm 2}\rangle$ (Eq.(5a), general factor is not important) shifted by $\alpha_1+\alpha_{In}$, provided that $a_{\pm 1}=(\alpha_1^*+2\alpha_{In}^*)/(\alpha_{In}^*(\alpha_1^*+\alpha_{In}^*))$ and $a_{\pm 2}=\sqrt{2}/(\alpha_{In}^*(\alpha_1^*+\alpha_{In}^*))$. Thus, optical scheme in Fig. 1(a) results in either

$$D(\alpha_{In})|\Psi_{\pm 1}\rangle, \quad (6a)$$

if we use single photon addition $N=1$ or

$$D(\alpha_{In}+\alpha_1)|\Psi_{\pm 2}\rangle, \quad (6b)$$

if we use two single photon additions $N=2$ with one intermediate displacement operator with an amplitude $\alpha_1$.

The fidelity between arbitrary states $F=|\langle\varphi_t|\varphi\rangle|$ is a measure of how close a state $|\varphi\rangle$ to the target state $|\varphi_t\rangle$. It is unity when the two states are identical, while it is zero when the two states are orthogonal to each other. In application to our case, $|\varphi\rangle$ can be either states (6a) or (6b), while $|\varphi_t\rangle$ is displaced squeezed SCSs defined in general case as

$$|DSSCS_\pm(\alpha_{SCS},\alpha,r)\rangle =$$
$$D(\alpha)S(r)|SCS_\pm(\alpha_{SCS})\rangle = \quad (7)$$
$$D(\alpha)S(r)N_\pm(\alpha_{SCS})$$
$$(|0,\alpha_{SCS}\rangle \pm |0,-\alpha_{SCS}\rangle)$$

where $S(r)=\exp\left(\frac{r}{2}(a^{+2}-a^2)\right)$ is a squeezing operator with $r$ being a squeezing parameter [31]. If we take $r=\alpha=0$, we deal with regular SCSs (Eq. (1)), if we choose $\alpha=0$ and $r\ne 0$ then we have squeezed SCSs [21-23] and if we use $r=0$ and $\alpha\ne 0$, we obtain displaced SCSs. Then, fidelity can be rewritten as

$$F_{\pm N} =$$
$$\left|\left\langle\begin{array}{c}SCS_\pm(\alpha_{SCS})S^+(r) \\ D^+(\alpha)D(\beta_N)\end{array}\middle|\Psi_{\pm N}\right\rangle\right|^2, \quad (8a)$$

where $\beta_N$ is either $\beta_1=\alpha_{In}$ (Eq. (6a)) or $\beta_2=\alpha_{In}+\alpha_1$ (Eq. (6b)). The formula (8a) can be adjusted for our cases as

$$F_{\pm 1} =$$
$$\left|\left\langle\begin{array}{c}SCS_\pm(\alpha_{SCS})S^+(r) \\ D^+(\alpha)D(\alpha_{In})\end{array}\middle|\Psi_{\pm 1}\right\rangle\right|^2$$
$$= \left|\left\langle\begin{array}{c}SCS_\pm(\alpha_{SCS}) \\ D(\gamma_{\pm 1})S(-r)\end{array}\middle|\Psi_{\pm 1}\right\rangle\right|^2 = \quad (8b)$$
$$\left|\langle SCS_\pm(\alpha_{SCS},\gamma_{\pm 1})S(-r)|\Psi_{\pm 1}\rangle\right|^2$$

$$F_{\pm 2} =$$
$$\left|\left\langle\begin{array}{c}SCS_\pm(\alpha_{SCS})S^+(r_\pm) \\ D^+(\alpha_\pm)D(\alpha_1+\alpha_{In})\end{array}\middle|\Psi_{\pm 2}\right\rangle\right|^2 =$$
$$\left|\left\langle\begin{array}{c}SCS_\pm(\alpha_{SCS}) \\ D(\gamma_{\pm 2})S(-r_\pm)\end{array}\middle|\Psi_{\pm 2}\right\rangle\right|^2 = \quad (8c)$$
$$\left|\langle SCS_\pm(\alpha_{SCS},\gamma_{\pm 2})S(-r_\pm)|\Psi_{\pm 2}\rangle\right|^2$$

where
$$\gamma_{\pm 1}=\cosh r(\alpha_{In}-\alpha)-\sinh r(\alpha_{In}-\alpha)^*,$$

$\gamma_{\pm 2} = \cosh r_{\pm} (\alpha_{In} + \alpha_1 - \alpha_{\pm}) - \sinh r_{\pm} (\alpha_{In} + \alpha_1 - \alpha_{\pm})^*$ , and $|SCS_{\pm}(\alpha_{SCS}, \gamma_{\pm 1})\rangle$, $|SCS_{\pm}(\alpha_{SCS}, \gamma_{\pm 2})\rangle$ are $\gamma_{\pm 1}-$ and $\gamma_{\pm 2}-$ representations of the regular even/odd SCSs, respectively, (Eq. (1), where $Q = \pm \pi/4$ is chosen). Choice of the input conditions may be determined by the aims. The directions of the developments for the generation of SCSs may be the following. Sometimes, one needs to generate SCSs with larger amplitude $\alpha_{SCS} \geq 2$ for macroscopic tests of quantum theory. For quantum information processing, it is important to construct SCSs with higher fidelities $F > 0.99$. Ideal case is to look for optimal conditions needed to generate SCSs with larger amplitudes and higher fidelities.

Further analysis involves both numerical and analytical methods. Indeed, it is possible numerically to look for the parameters $\gamma_{\pm 1}$, $a_{\pm 1}$, and $r$ under which the fidelity $F_{\pm 1}$ (8b) takes maximal values. Knowing the parameters $\gamma_{\pm 1}$, $a_{\pm 1}$, and $r_{\pm}$, we can calculate analytically $\alpha_{In}$, $\alpha_{\pm}$ from their definition. The calculated parameters that have to be used in optical scheme in Fig 1(a) are collected in Table 1.

| $\alpha_{SCS}$ | $Q = \dfrac{\pi}{4}$ | $Q = -\dfrac{\pi}{4}$ |
|---|---|---|
| 0.8 | $F_{+1} = 0.988095$, $r_+ = -0.349324$<br>$\alpha_{In} = i1.83218$,<br>$\alpha_+ = i2.26764$ | $F_{-1} = 0.999376$, $r_- = -0.207344$<br>$\alpha_{In} = 0$,<br>$\alpha_- = 0$ |
| 0.9 | $F_{+1} = 0.97744$, $r_+ = -0.397861$<br>$\alpha_{In} = i1.49474$,<br>$\alpha_+ = i1.98742$ | $F_{-1} = 0.998584$, $r_- = -0.258353$<br>$\alpha_{In} = 0$,<br>$\alpha_- = 0$ |
| 1 | $F_{+1} = 0.962444$, $r_+ = -0.445031$<br>$\alpha_{In} = i1.2464$,<br>$\alpha_+ = i1.78867$ | $F_{-1} = 0.997109$, $r = -0.31257$<br>$\alpha_{In} = 0$,<br>$\alpha_- = 0$ |
| 1.1 | $F_{+1} = 0.943626$, $r_+ = -0.491368$<br>$\alpha_{In} = i1.05247$,<br>$\alpha_+ = i1.6373$ | $F_{-1} = 0.994411$, $r_- = -0.36893$<br>$\alpha_{In} = 0$,<br>$\alpha_- = 0$ |
| 1.2 | $F_{+1} = 0.922092$, $r_+ = -0.537234$<br>$\alpha_{In} = i0.900828$,<br>$\alpha_+ = i1.52201$ | $F_{-1} = 0.990085$, $r_- = -0.426398$<br>$\alpha_{In} = 0$,<br>$\alpha_- = 0$ |

**Table 1** Values of initial coherent seed $\alpha_{In}$ in Fig. 1(a) with single photon addition under which the output approximates either even DSSCS $D(\alpha_+)S(r_+)|SCS_+(\alpha_{SCS})\rangle$ or odd DSSCS $D(\alpha_-)S(r_-)|SCS_-(\alpha_{SCS})\rangle$ with maximal fidelity.

The same approach combining both numerical and analytical tools can be used for the case with two photon additions $N=2$ and one intermediate displacement operator between them. It is possible numerically to look for parameters $a_{\pm 1}$, $a_{\pm 2}$, $\gamma_{\pm 2}$ and $r_\pm$ under which the fidelity (8c) takes maximal value. It enables to estimate parameters $\alpha_{In}$, $\alpha_1$ and $\alpha_\pm$ for the optical scheme in Fig. 1(a) as

$$\alpha_{In} = \pm i|\alpha_{In}| = \pm i\sqrt{\frac{\sqrt{2}}{a_{+2}}}, \quad (9a)$$

$$\alpha_1 = \mu 2i|\alpha_{In}|, \quad (9b)$$

$$\alpha_+ = \mu i|\alpha_{In}|, \quad (9c)$$

for even SCS $(Q = \pi/4)$, where $a_{+2} > 0$ and $a_{+1} = 0$ [21] and

$$\alpha_{In}^* = \frac{a_{-1}}{\sqrt{2}a_{-2}} \pm \frac{\sqrt{D}}{2}, \quad (10a)$$

$$\alpha_1^* = \mu\sqrt{D}, \quad (10b)$$

$$D = 2\left(\frac{a_{-1}}{a_{-2}}\right)^2 - \frac{4\sqrt{2}}{a_{-2}}, \quad (10c)$$

for odd SCS $(Q = \pi/4)$, while an amplitude of shift $\alpha_-$ follows from definition of $\gamma_{-2}$. Corresponding parameters $\alpha_{In}$, $\alpha_1$, $\alpha_\pm$ and $r_\pm$ that have to be used in optical scheme in Fig. 1(a) with two photon additions and one intermediate displacement operator to reach maximal fidelity with squeezed SCSs are collected in Table 2.

| $\alpha_{SCS}$ | $Q = \dfrac{\pi}{4}$ | $Q = -\dfrac{\pi}{4}$ |
|---|---|---|
| 1 | $F_{+2} = 0.9999$, $r_+ = -0.179612$<br><br>a) $\alpha_{In} = i1.5904$, $\alpha_1 = -i3.1808$, $\alpha_+ = -i1.5904$<br><br>b) $\alpha_{In} = -i1.5904$, $\alpha_1 = i3.1808$, $\alpha_+ = i1.5904$ | $F_{-2} = 0.997473$, $r_- = -0.253791$<br><br>a) $\alpha_{In} = i0.243421$, $\alpha_1 = -i4.09883$, $\alpha_- = -i4.09884$<br><br>b) $\alpha_{In} = -i3.85488$, $\alpha_1 = i4.09883$, $\alpha_- = 0$ |
| 1.1 | $F_{+2} = 0.999738$, $r_+ = -0.215319$<br><br>a) $\alpha_{In} = i1.45591$, $\alpha_1 = -i2.91183$, $\alpha_+ = -i1.45591$<br><br>b) $\alpha_{In} = -i1.45591$, $\alpha_1 = i2.91183$, $\alpha_+ = i1.45591$ | $F_{-2} = 0.995285$, $r_- = -0.292058$<br><br>a) $\alpha_{In} = i0.279104$, $\alpha_1 = -i3.56808$, $\alpha_- = -i3.56809$<br><br>b) $\alpha_{In} = -i3.28898$, $\alpha_1 = i3.56808$, $\alpha_- = 0$ |
| 1.2 | $F_{+2} = 0.999392$, $r_+ = -0.253272$<br><br>a) $\alpha_{In} = i1.34629$, $\alpha_1 = -i2.69259$, | $F_{-2} = 0.991945$, $r_- = -0.330436$<br><br>a) $\alpha_{In} = i0.312901$, $\alpha_1 = -i3.17489$, |

|  | $\alpha_+ = -i1.34629$ | $\alpha_- = -i3.17491$ |
|---|---|---|
|  | b) $\alpha_{In} = -i1.34629$, $\alpha_1 = i2.69259$, $\alpha_+ = i1.34629$ | b) $\alpha_{In} = -i2.86199$, $\alpha_1 = i3.17489$, $\alpha_- = 0$ |
| 1.3 | $F_{+2} = 0.998728$, $r_+ = -0.293054$  a) $\alpha_{In} = i1.25598$, $\alpha_1 = -i2.251196$, $\alpha_+ = -i1.25598$  b) $\alpha_{In} = -i1.25598$, $\alpha_1 = i2.51196$, $\alpha_+ = i1.25598$ | $F_{-2} = 0.987245$, $r_- = -0.368812$  a) $\alpha_{In} = i0.344249$, $\alpha_1 = -i2.87582$, $\alpha_- = -i2.87586$  b) $\alpha_{In} = -i2.53147$, $\alpha_1 = i2.87582$, $\alpha_- = 0$ |
| 1.4 | $F_{+2} = 0.997583$, $r_+ = -0.334228$  a) $\alpha_{In} = i1.18095$, $\alpha_1 = -i2.3619$, $\alpha_+ = -i1.18095$  b) $\alpha_{In} = -i1.18095$, $\alpha_1 = i2.3619$, $\alpha_+ = i1.19095$ | $F_{-2} = 0.981078$, $r_- = -0.407125$  a) $\alpha_{In} = i0.373226$, $\alpha_1 = -i2.64328$, $\alpha_- = -i2.64334$  b) $\alpha_{In} = -i2.27005$, $\alpha_1 = i2.64328$, $\alpha_- = 0$ |
| 1.5 | $F_{+2} = 0.995765$, $r_+ = -0.376383$  a) $\alpha_{In} = i1.11822$, $\alpha_1 = -i2.23643$, $\alpha_+ = -i1.11822$  b) $\alpha_{In} = -i1.11822$, $\alpha_1 = i2.23643$, $\alpha_+ = i1.11822$ | $F_{-2} = 0.987245$, $r_- = -0.445339$  a) $\alpha_{In} = i0.399473$, $\alpha_1 = -i2.45894$, $\alpha_- = -i2.45903$  b) $\alpha_{In} = -i2.05947$, $\alpha_1 = i2.45894$, $\alpha_- = 0$ |
| 1.6 | $F_{+2} = 0.993085$, $r_+ = -0.419055$  a) $\alpha_{In} = i1.06794$, $\alpha_1 = -i2.13588$, $\alpha_+ = -i1.06794$  b) $\alpha_{In} = -i1.06794$, $\alpha_1 = i2.13588$, $\alpha_+ = i1.06794$ | $F_{-2} = 0.964491$, $r_- = -0.483419$  a) $\alpha_{In} = i0.423166$, $\alpha_1 = -i2.31033$, $\alpha_- = -i2.31047$  b) $\alpha_{In} = -i1.88716$, $\alpha_1 = i2.31033$, $\alpha_- = 0$ |
| 1.7 | $F_{+2} = 0.989373$, $r_+ = -0.46194597$  a) $\alpha_{In} = i1.02351$, $\alpha_1 = -i2.04701$, $\alpha_+ = -i1.02351$  b) $\alpha_{In} = -i1.02351$, $\alpha_1 = i2.04701$, $\alpha_+ = i1.02351$ | $F_{-2} = 0.954387$, $r_- = -0.521336$  a) $\alpha_{In} = i0.444419$, $\alpha_1 = -i2.18895$, $\alpha_- = -i2.18914$  b) $\alpha_{In} = -i1.74453$, $\alpha_1 = i2.18895$, $\alpha_- = 0$ |

**Table 2** Values of initial coherent seed $\alpha_{In}$ and intermediate displacement operator $\alpha_1$ in Fig. 1(a) under which the output approximates either even DSSCS $D(\alpha_+)S(r_+)|SCS_+(\alpha_{SCS})\rangle$ or odd DSSCS $D(\alpha_-)S(r_-)|SCS_-(\alpha_{SCS})\rangle$ with maximal fidelity.

Comparing the data collected in Tables 1 and 2, it is possible to make the following conclusions. So, $\alpha_{In} \approx 0$ is chosen (right column in Table 1), this means that vacuum as input is used to generate odd SCS in optical scheme in Fig. 1(a). As $\alpha \approx 0$ is taken (right column in Table 1), it means that such output approximates odd squeezed SCS (not displaced) that is total agreement with previous results [15]. Left column in Table 1 shows that seed coherent state is used to generate displaced by $\alpha_+ \neq 0$ even squeezed SCS with a little less fidelity than one in the case of odd SSCS generation. Since only one photon creation operator $a^+$ is used to generate single photon added coherent state (SPACS), the method may look attractive for generation of the SCSs due to its simplicity. So, the SPACS has been experimentally demonstrated in [31]. Comparing results of [31] with those presented in Table 1, it is possible to claim the SPACSs generated in [31] do not approximate DSSCSs as amplitudes of experimental seed coherent states were chosen out of range needed for generation of the DSSCSs. As can be seen from Table 2, it is possible to increase amplitudes of generated SSCSs in optical scheme in Fig. 1(a) with two single photon additions with intermediate displacement operator with an amplitude $\alpha_1$ between them. It follows from Table 2 that input coherent state with both amplitude $i|\alpha_{In}|$ and $-i|\alpha_{In}|$ can be used. Then, output is even/odd SSCS in dependency on choice of $\alpha_{In}$ shifted relative each other by a value $\alpha_+ - \alpha_-$.

### III. HADAMATD GATE FOR COHERENT STATES

It is well known there is no fundamental reason to restrict oneself to physical systems with two-dimensional Hilbert spaces for the encoding. It may be more natural in some contexts to encode logical states as a superposition over a large number of basic states as it occurs in quantum computation with coherent optical states, where two coherent states with amplitudes differing by $\pi$ $|0,\pm\alpha\rangle$ are used as computational basis [10,11]. So, we can define a local operation $R(Q)$ as

$$R(Q)|0,\alpha\rangle = \cos Q|0,\alpha\rangle + \sin Q|0,-\alpha\rangle, \quad (11a)$$

$$R(Q)|0,-\alpha\rangle = \sin Q|0,\alpha\rangle - \cos Q|0,-\alpha\rangle, \quad (11b)$$

which is nonunitary due to the nonorthogonality of $|0,\alpha\rangle$ and $|0,-\alpha\rangle$. However, $R(Q)$ becomes approximately unitary when the overlap between the two coherent states, $\langle 0,\alpha|0,-\alpha\rangle = \exp(-2|\alpha|^2)$, approaches zero. It should be noted that this overlap goes rapidly to zero as $\alpha$ increases. If we take $Q = \pi/4$, the local operation $R(Q)$ becomes Hadamard gate that transforms $|0,\alpha\rangle$ to the state

$$R(Q=\pi/4)|0,\alpha\rangle = \left(\begin{array}{c}|0,\alpha\rangle + \\ |0,-\alpha\rangle\end{array}\right)\Big/\sqrt{2}, \quad (11c)$$

and $|0,-\alpha\rangle$ to the state

$$R(Q=\pi/4)|0,-\alpha\rangle = \left(\begin{array}{c}|0,\alpha\rangle - \\ |0,-\alpha\rangle\end{array}\right)\Big/\sqrt{2}. \quad (11d)$$

Hadamard gate is a mainframe elementary quantum gate used for performance of quantum tasks with coherent states. To achieve arbitrary $1-$ bit rotation, we need operate $U(\pi/4)$ and $U(-\pi/4)$ which are rotations by $\pi/2$ and $-\pi/2$, respectively, around the $x$ axis. The unitary operations $U(\pi/4)$ and $U(-\pi/4)$ can be realized using a Kerr nonlinear interaction [1,3]. The interaction Hamiltonian of a single-mode Kerr nonlinearity is $H_{NL} = \eta\Omega(a^+ a)^2$, where $\Omega$ is the strength of the Kerr nonlinearity. When the interaction time $t$ in the medium is $\pi/\Omega$ coherent states evolve as (11c) and (11d) up to relative phase shift by $\pi/2$. Optical fiber is well known example of a medium with Kerr nonlinearity but only statistical mixing of the states $|0,\alpha\rangle$ and $|0,-\alpha\rangle$ occurs on output of long fiber instead of (11c) and (11d), respectively, due to decoherence effects when optical beams propagate inside the fiber. It may be main drawback for development of the quantum protocols with coherent optical states. For a universal gate operation, a CNOT gate is required besides 1– bit rotation. It was found the CNOT operation can be realized using a teleportation protocol. Instead of the local regular transformations (11(a)-11(d)) it is possible to consider the squeezed rotations of the coherent states as

$$SR(Q)|0,\alpha\rangle = S(r)\begin{pmatrix}\cos Q|0,\alpha\rangle + \\ \sin Q|0,-\alpha\rangle\end{pmatrix}, \quad (12a)$$

$$SR(Q)|0,-\alpha\rangle = S(r)\begin{pmatrix}\sin Q|0,\alpha\rangle - \\ \cos Q|0,-\alpha\rangle\end{pmatrix}, \quad (12b)$$

in general case of arbitrary value of $Q$ and

$$SR(Q)|0,\alpha\rangle = S(r)(\cos Q|0,\alpha\rangle + \sin Q|0,-\alpha\rangle)/\sqrt{2}, \quad (12c)$$

$$SR(Q)|0,-\alpha\rangle = S(r)(\sin Q|0,\alpha\rangle - \cos Q|0,-\alpha\rangle)/\sqrt{2}, \quad (12d)$$

in partial case of $Q = \pi/4$ (Hadamard gate). Applying final squeezing operator $S(-r)$ to the formulas (12a)-(12d), we obtain regular rotations (Eqs. (11a)-(11d)).

Analyzing the data in Tables 1 and 2, we propose to make use of optical scheme in Fig. 1(b) to use it for the following nonunitary transformations with input coherent states (Eqs. (1,2)) with amplitudes $\pm\alpha_Q$

$$|0,\alpha_Q\rangle \to$$
$$D(\alpha_+)S(r)SCS_Q(\alpha_{SCS}) \approx \quad , \quad (13a)$$
$$D(\alpha_+)S(r)\begin{pmatrix}\cos Q|0,\alpha_{SCS}\rangle + \\ \sin Q|0,-\alpha_{SCS}\rangle\end{pmatrix}$$

$$|0,-\alpha_Q\rangle \to$$
$$D(\alpha_-)S(r)SCS_{Q-\pi/2}(\alpha_{SCS}) \approx , \quad (13b)$$
$$D(\alpha_-)S(r)\begin{pmatrix}\sin Q|0,\alpha_{SCS}\rangle - \\ \cos Q|0,-\alpha_{SCS}\rangle\end{pmatrix}$$

in general case and with amplitudes $\pm\alpha_{GH}$

$$|0,\alpha_{GH}\rangle \to$$
$$D(\alpha_+)S(r)SCS_+(\alpha_{SCS}) \approx \quad , \quad (13c)$$
$$D(\alpha_+)S(r)\begin{pmatrix}|0,\alpha_{SCS}\rangle + \\ |0,-\alpha_{SCS}\rangle\end{pmatrix}/\sqrt{2}$$

$$|0,-\alpha_{GH}\rangle \to$$
$$D(\alpha_-)S(r)SCS_-(\alpha_{SCS}) \approx \quad , \quad (13d)$$
$$D(\alpha_-)S(r)\begin{pmatrix}|0,\alpha_{SCS}\rangle - \\ |0,-\alpha_{SCS}\rangle\end{pmatrix}/\sqrt{2}$$

in the case of Hadamard gate. It is worth noting that it may happen that $\alpha_Q \ne \alpha_{SCS}$ and $\alpha_{GH} \ne \alpha_{SCS}$ in expressions (13a)-(13d). We are more interested in investigation of optical scheme whose outcome approximates squeezed either even or odd SCSs of larger amplitude ($N = 2$) shifted relative each other in general case as shown in formulas (13c) and (13d). Therefore, the optical scheme in Fig. 1(b) starts with a coherent state with either an amplitude $\alpha_{GH}$ or $-\alpha_{GH}$. After that displacement operator with an amplitude $\beta$ is applied to provide

$$D(\beta)|0,\alpha_{HG}\rangle = \exp(i\varphi)D(\alpha_{GH}+\beta)|0\rangle = \exp(i\varphi)|0,\alpha_{HG}+\beta\rangle = \exp(i\varphi)|0,\alpha_{In}\rangle$$

for the generation of the squeezed even SCS and

$$D(\beta)|0,-\alpha_{HG}\rangle = \exp(i\varphi)D(\alpha_{GH}+\beta)|0\rangle =$$
$$\exp(i\varphi)|0,\alpha_{HG}+\beta\rangle = \exp(i\varphi)|0,\alpha_{In}\rangle$$

for the generation of the squeezed odd SCS. Analysis of Table 2 shows it is possible to choose shifting amplitude $\alpha_1$ of intermediate displacement operator for the optical scheme in Fig. 1(b) equal for both even and odd SSCSs construction and, moreover, to provide the condition $r_+ = r_- = r$ for output states. Thus, optical scheme in Fig. 1(b) results in outcome that is described by expressions (13c) and (13d) provided that only amplitude of the input coherent state $\alpha_{GH}$ changes on opposite $-\alpha_{GH}$ and vise versa. We can also start with either $\alpha_{SCS}$ or $-\alpha_{SCS}$ to provide equality of input and output amplitudes of the generated SSCSs in optical scheme in Fig. 1(b). To do it we can additionally supply the optical scheme in Fig. 1(b) by phase shifter by $\pi/2$ and absorbing medium (dashed rectangle on figure) to provide $|0,\alpha_{SCS}\rangle \to |0,i\alpha_{SCS}\rangle \to |0,i\alpha_{SCS}\exp(-\Gamma)\rangle = |0,\alpha_{HG}\rangle$ and

$$|0,-\alpha_{SCS}\rangle \to |0,-i\alpha_{SCS}\rangle \to$$
$$|0,-i\alpha_{SCS}\exp(-\Gamma)\rangle = |0,-\alpha_{HG}\rangle,$$

respectively, where $\Gamma$ is absorbing factor of the medium. Then, we provide equality of input and output amplitudes of SSCSs due to amplification action of the squeezing operator. We collect all optimal data in Table 3 that are used for the construction of the Hadamard gate with coherent states

| $\alpha_{SCS}$, $r$, $\alpha_1$, $\alpha_{HG}$, $\beta$ | $Q = \dfrac{\pi}{4}$ | $Q = -\dfrac{\pi}{4}$ |
|---|---|---|
| $\alpha_{SCS} = 1.3$, $r = -0.351$ <br><br> a) $\alpha_1 = -i2.87582$, $\alpha_{HG} = i0.546781$, $\beta = i0.89113$ <br><br> b) $\alpha_1 = i2.87582$, $\alpha_{HG} = i0.54678$, $\beta = -i1.98469$ | $F_{+2} = 0.986582$ <br><br> a) $\alpha_{In} = i1.43791$, $\alpha_+ = -i1.43791$ <br><br><br> b) $\alpha_{In} = -i1.43791$, $\alpha_+ = i1.43791$ | $F_{-2} = 0.986539$ <br><br> a) $\alpha_{In} = i0.344349$, $\alpha_- = -i2.87586$ <br><br><br> b) $\alpha_{In} = -i2.53147$, $\alpha_- = 0$ |
| $\alpha_{SCS} = 1.4$, $r = -0.40712$ <br><br> a) $\alpha_1 = -i2.64328$, $\alpha_{HG} = i0.474207$, $\beta = i0.847433$ <br><br> b) $\alpha_1 = i2.64328$, $\alpha_{HG} = i0.474205$, $\beta = -i1.79585$ | $F_{+2} = 0.986162$ <br><br> a) $\alpha_{In} = i1.32164$, $\alpha_+ = -i1.32164$ <br><br><br> b) $\alpha_{In} = -i1.32164$, $\alpha_+ = i1.32164$ | $F_{-2} = 0.981078$ <br><br> a) $\alpha_{In} = i0.373226$, $\alpha_- = -i2.64334$ <br><br><br> b) $\alpha_{In} = -i2.27005$, $\alpha_- = 0$ |
| $\alpha_{SCS}=1.5, r=-0.445339$ | $F_{+2}=0.985525$ | $F_{-2}=0.973453$ |
| a) $\alpha_1=-i2.45894$, $\alpha_{HG}=i0.414998$, $\beta=i0.8144715$ | a) $\alpha_{In}=i1.22947$, $\alpha_+=-i1.22947$ | a) $\alpha_{In}=i0.399473$, $\alpha_-=-i2.45903$ |
| b) $\alpha_1=i2.45894$, $\alpha_{HG}=i0.415$, $\beta=-i1.64447$ | b) $\alpha_{In}=-i1.22947$, $\alpha_+=i1.22947$ | b) $\alpha_{In}=-i2.05947$, $\alpha_-=0$ |
| $\alpha_{SCS}=1.6, r=-0.483418$ | $F_{+2}=0.983888$ | $F_{-2}=0.964491$ |
| a) $\alpha_1=-i2.31033$, $\alpha_{HG}=i0.366002$, $\beta=i0.789168$ | a) $\alpha_{In}=i1.15517$, $\alpha_+=-i1.15517$ | a) $\alpha_{In}=i0.423166$, $\alpha_-=-i2.31047$ |
| b) $\alpha_1=i2.31033$, $\alpha_{HG}=i0.365995$, $\beta=-i1.52116$ | b) $\alpha_{In}=-i1.15517$, $\alpha_+=i1.15517$ | b) $\alpha_{In}=-i1.88716$, $\alpha_-=0$ |
| $\alpha_{SCS}=1.7, r=-0.521336$ | $F_{+2}=0.98118$ | $F_{-2}=0.954387$ |
| a) $\alpha_1=-2.188915$, $\alpha_{HG}=i0.32503$, $\beta=i0.769449$ | a) $\alpha_{In}=i1.09448$, $\alpha_+=-i1.09448$ | a) $\alpha_{In}=i0.444419$, $\alpha_-=-i2.18914$ |
| b) $\alpha_1=i2.188915$, $\alpha_{HG}=i0.325025$, $\beta=-i1.41951$ | b) $\alpha_{In}=-i1.09448$, $\alpha_+=i1.09448$ | b) $\alpha_{In}=-i1.74453$, $\alpha_-=0$ |

**Table 3** Values of the parameters used in optical scheme Fig. 1(b) to provide it as Hadamard gate with coherent states.

Thus, optical scheme in Fig. 1(b) works as displacing Hadamard gate since output states are the squeezed SCSs in dependency on input basic states either $|0,\alpha_{GH}\rangle$ ($|0,\alpha_{SCS}\rangle$) or $|0,-\alpha_{GH}\rangle$ ($|0,-\alpha_{SCS}\rangle$) shifted relative each by a value $\alpha_+ - \alpha_-$. Some examples of Wigner functions of the output states of the optical scheme in Fig. 1(b) and the states that they approximate are shown in Fig. 2(a-d) and 3(a-d). All the parameters for the construction of Wigner functions are taken from Table 3. Interference features of the states manifest in $p-$ distribution while separated peaks are observed in $x-$ distribution of generated states and squeezed even/odd SCSs. The states shown in figures 2(a), 2(c) approximate even/odd SSCSs (Fig. 2(b), 2(d)) with amplitude $\alpha_{SCS}=1.4$, respectively. The states shown in figures 3(a), 3(c) approximate even/odd SSCSs (Fig. 3(b), 3(d)) with amplitude

$\alpha_{SCS} = 1.5$, respectively. As can been seen from figures 2,3, shift of the squeezed even/odd SCSs occurs only in $p-$ distribution. The shift between squeezed even and odd SCSs on phase plane follows from data given in Table 3. So, for example, the shift between generated states on phase plane is $|\alpha_+ - \alpha_-| = 1.32164$ for $\alpha_{SCS} = 1.4$ (Fig. 2(a-d)) and $|\alpha_+ - \alpha_-| = 1.22947$ for $\alpha_{SCS} = 1.5$ (Fig. 3(a-d)).

Our approach is based on use of single photon additions. It is well known the single photon addition can be obtained probabilistically with help of parametric down converter. Probability of such an event is low. Nevertheless, SPACS was experimentally generated in [32] and probability to register only one photon at ancillary mode on output of the down converter prevails over the probabilities to register more than one photon. It can mean that problem of resolving number states becomes negligible and, as consequence, we may use silicon avalanche photodiodes operating at the visible wavelength having relatively high efficiency and a small dark count rate. If the dark count rate of photodetector is negligible, then output state may be in a mixed state represented as $(1-P)W_{SPACS}(\alpha) + PW_0(\alpha)$, where $W_{SPACS}(\alpha)$ is the Wigner function of the SPACS and $W_0(\alpha)$ is the Wigner function of vacuum and $P$ is the probability to register occasional photon. Construction of higher-order states $|\Psi_{\pm 2}\rangle$ requires intermediate displacement operator and additional single-photon addition that decreases the success probability of the device in Fig. 1(a). A displacement operator $D(\beta)$ with amplitude $\beta$ can be approximated by beam splitter with high transmittivity $T \to 1$ mixing input filed with ancillary strong coherent field $|0,\xi\rangle$ ($\xi \gg 1$). Then, output can be evaluated as $(1-P)W_{\pm 2}(\alpha) + PW_\alpha(\alpha)$, where $W_{\pm 2}(\alpha)$ is the Wigner function of either $|\Psi_{\pm 2}\rangle$ and $W_\alpha(\alpha)$ is Wigner function of coherent state, if we neglect probability to register two occasional photons. Thus, the fidelity of the generated states in practice depends on dark count rate and success probability of the method drops with increase of $N$.

IV. CONCLUSION

It is well known it is hardly possible to generate regular SCSs with large amplitude with current level of technologies. It is natural way to overcome the drawback is to approximate regular SCSs to any degree of accuracy by some states involving either finite in infinite number of terms. The problem of arbitrary rotations of coherent states is extremely challenging. Our approach proposes to resolve the problem. To do it we made use of possibility to decompose any arbitrary one-mode state, in particular the SCSs, in terms of the displaced number states with arbitrary amplitudes (so called $\alpha-$ representation). This decomposition enables to show type of generated SCSs (even or odd) does not depend on photon parity. Photon parity can be defined only for the SCSs at $0-$ representation. The main motivation to use the representation is to consider problems of rotations of coherent states in application them to quantum computation. We used a method developed in [30], as it is shown in Fig. 1(a), 1(b), to construct the states that approximate displaced squeezed SCSs and applied it to make Hadamard gate with coherent states. The ability to investigate the elementary actions of the bosonic creation operators on a seed coherent state is of interest both as a tool to take closer look at fundamental events in quantum physics and natural extensions toward exotic quantum entities, such as SCSs. Our analysis shows that it is possible to choose parameters of the optical scheme in Fig. 1(a), 1(b) in such a way the output becomes sensitive to seed coherent state that allows for one to construct local rotations of qubits, in particular Hadamard gate, consisting of coherent states with high fidelity. One should note it is not rotations as they are defined by expressions (11a) and (11b). The outcomes are the squeezed SCSs shifted relative each other by some quantity along $p$ – axis. The Hadamard gate which effects transformation as in Fig. 1(b) cannot be unitary. Possible use of the Hadamard gate for quantum computation with coherent states deserves separate investigation [22]. All parameters needed to construct either even or odd DSSCSs in dependency on seed coherent states are presented in Table 1, 2, respectively, and, finally, range of parameters of the optical scheme in Fig. 1(b) which can be used for the construction of displacing Hadamard gate is given in Table 3. In the short-term, this approach extends set of the states that may be used for quantum information processing and add new methods for manipulations with coherent state qubits.

**Appendix A. Decomposition in terms of the displaced number states**

We are going to make use of the following representation of the coherent state

$$|0,\alpha_{SCS}\rangle = D(\alpha_{SCS})|0\rangle =$$
$$\exp(-|\alpha_{SCS}|^2/2)$$
$$\exp(\alpha_{SCS}a^+)\exp(\alpha_{SCS}a)|0\rangle =$$
$$\exp(-|\alpha_{SCS}|^2/2)\exp(\alpha_{SCS}a^+)|0\rangle =$$
$$\exp(-|\alpha_{SCS}|^2/2)\exp((\alpha+\beta)a^+)|0\rangle =$$
$$\exp(-(|\alpha_{SCS}|^2-|\alpha|^2)/2)\exp(\beta a^+)$$
$$\exp(-|\alpha|^2/2)\exp(\alpha a^+)|0\rangle =$$
$$\exp(-(|\alpha_{SCS}|^2-|\alpha|^2)/2)$$
$$\exp(\beta a^+)|0,\alpha\rangle =$$
$$\exp(-(|\alpha_{SCS}|^2-|\alpha|^2)/2)$$
$$\sum_{n=0}^{\infty}\frac{(\beta a^+)^n}{n!}|0,\alpha\rangle \qquad , \text{(A1)}$$

where $\alpha_{SCS} = \alpha + \beta$ $(\beta = \alpha_{SCS} - \alpha)$ and $\alpha$, $\beta$ are the arbitrarily chosen numbers. Consider $a^{+n}|0,\alpha\rangle$ using the well-known formulas [31]. Then, we have

$$a^{+n}|0,\alpha\rangle = D(\alpha)D^+(\alpha)a^{+n}D(\alpha)|0\rangle =$$
$$D(\alpha)(a^+ + \alpha^*)^n|0\rangle =$$
$$D(\alpha)\sum_{k=0}^{n}C_n^k\sqrt{k!}\alpha^{*n-k}|k\rangle = \qquad \text{. (A2)}$$
$$\sum_{k=0}^{n}\frac{n!\sqrt{k!}\alpha^{*n-k}}{k!(n-k)!}|k,\alpha\rangle$$

Using (A2), it is possible to transform $\sum_{n=0}^{\infty}((\beta a^+)^n/n!)|0,\alpha\rangle$ to

$$\sum_{n=0}^{\infty} \frac{(\beta a^+)^n}{n!} |0,\alpha\rangle = |0,\alpha\rangle +$$
$$\beta(|1,\alpha\rangle + \alpha^*) + (\beta^2/2!)$$
$$\begin{pmatrix} \sqrt{2}|2,\alpha\rangle + 2\alpha^*|1,\alpha\rangle + \\ \alpha^{*2}|0,\alpha\rangle \end{pmatrix} + ...$$
$$+ (\beta^n/n!)\sum_{k=0}^{n} C_n^k \sqrt{k!}\alpha^{*n-k}|k\rangle + ... =$$
$$\begin{pmatrix} 1+\beta\alpha^* + \beta^2\alpha^{*2}/2! + ... \\ +\beta^n\alpha^{*n}/n! + ... \end{pmatrix}|0,\alpha\rangle +$$
$$\beta\begin{pmatrix} 1+\beta\alpha^* + ... + \\ C_n^1 \beta^{n-1}\alpha^{*n-1}/n! + ... \end{pmatrix}|1,\alpha\rangle +$$
$$\frac{\beta^2}{\sqrt{2!}}\begin{pmatrix} 1+ \\ \beta\alpha^* + ... + \\ C_n^2 \beta^{n-2}\alpha^{*n-2} 2!/n! + ... \end{pmatrix}|2,\alpha\rangle + ...$$
$$\frac{\beta^l}{\sqrt{l!}}\begin{pmatrix} 1+...+ \\ C_n^l \beta^{n-l}\alpha^{*n-l} l!/n! + ... \end{pmatrix}|l,\alpha\rangle + ... =$$
$$\begin{pmatrix} 1+\beta\alpha^* + \beta^2\alpha^{*2}/2! + ... \\ +\beta^n\alpha^{*n}/n! + ... \end{pmatrix}|0,\alpha\rangle +$$
$$\beta\begin{pmatrix} 1+\beta\alpha^* + ... \\ + \beta^{n-1}\alpha^{*n-1}/(n-1)! + ... \end{pmatrix}|1,\alpha\rangle +$$
$$\frac{\beta^2}{\sqrt{2!}}\begin{pmatrix} 1+\beta\alpha^* + ... + \\ \beta^{n-2}\alpha^{*n-2}/(n-2)! + ... \end{pmatrix}|2,\alpha\rangle + ...$$
$$\frac{\beta^l}{\sqrt{l!}}\begin{pmatrix} 1+...+ \\ \beta^{n-l}\alpha^{*n-l}/(n-l)! + ... \end{pmatrix}|l,\alpha\rangle + ... =$$
$$\exp(\beta\alpha^*)\sum_{l=0}^{\infty} \frac{\beta^l}{\sqrt{l!}}|l,\alpha\rangle \qquad (A3)$$

Finally, we have

$$|0,\alpha_{SCS}\rangle = \exp(-(|\alpha_{SCS}|^2 - |\alpha|^2)/2)$$
$$\exp(\beta\alpha^*)\sum_{l=0}^{\infty} \frac{\beta^l}{\sqrt{l!}}|l,\alpha\rangle =$$
$$\exp(-(\alpha_{SCS}^2 + |\alpha|^2)/2)$$
$$\exp(\alpha_{SCS}\alpha^*)\sum_{l=0}^{\infty} \frac{(\alpha_{SCS}-\alpha)^l}{\sqrt{l!}}|l,\alpha\rangle = \qquad (A4)$$
$$\exp(-(\alpha_{SCS}^2 + |\alpha|^2)/2)$$
$$\exp(\alpha_{SCS}\alpha^*)\sum_{l=0}^{\infty} \frac{\alpha_{SCS}^l}{\sqrt{l!}}\left(1-\frac{\alpha}{\alpha_{SCS}}\right)^l|l,\alpha\rangle$$

The same is applicable to the state $|0,-\alpha_{SCS}\rangle$ as

$$|0,-\alpha_{SCS}\rangle = \exp(-(\alpha_{SCS}^2 + |\alpha|^2)/2)$$
$$\exp(-\alpha_{SCS}\alpha^*) \qquad (A5)$$
$$\sum_{l=0}^{\infty} \frac{\alpha_{SCS}^l}{\sqrt{l!}}(-1)^l\left(1+\frac{\alpha}{\alpha_{SCS}}\right)^l|l,\alpha\rangle$$

So, if we take $\alpha = 0$ in (A4), then we have the following decomposition of the vacuum state in the basis of displaced number states

$$|0\rangle = |0,\alpha_{SCS} = 0\rangle = \exp(-|\alpha|^2/2)$$
$$\sum_{l=0}^{\infty} (-1)^l \frac{\alpha^l}{\sqrt{l!}}|l,\alpha\rangle \qquad (A6)$$

**Appendix B. Wigner functions of displaced squeezed SCSs and their approximations**

This Appendix concerns questions of the presentation of SCSs and their approximations on phase plane. So, Wigner functions of the even/odd SCSs can be expressed as

$$W_{\pm SCS}(\alpha) = N_\pm(\alpha_{SCS})$$
$$(W_0(\alpha) + W_{-0}(\alpha) \pm 2X_{\alpha_{SCS}}(\alpha)), \qquad (B1)$$

where $\alpha = x + ip$, $\alpha_{SCS} = x_{SCS} + ip_{SCS}$ and

$$W_0(\alpha) = \frac{2}{\pi}\exp\begin{pmatrix} -2(x-x_{SCS})^2 - \\ 2(p-p_{SCS})^2 \end{pmatrix}, \qquad (B2)$$

$$W_{-0}(\alpha) = \frac{2}{\pi}\exp\begin{pmatrix} -2(x+x_{SCS})^2 - \\ 2(p+p_{SCS})^2 \end{pmatrix}, \qquad (B3)$$

$$X_{\alpha_{SCS}}(\alpha) = \frac{2}{\pi}\exp(-2x^2 - 2p^2) \cos(4(xp_{SCS} - px_{SCS})) \quad (B4)$$

As we deal with displaced squeezed SCSs, we present their Wigner functions. If Wigner function of the SCSs is given by $W_{\pm SCS}(\alpha)$, then it is possible to show that Wigner function transforms

$$W_{\pm SCS}(\alpha) \to W_{\pm DSSCS}\begin{pmatrix}\cosh r(\alpha - \xi) - \\ \sinh r(\alpha^* - \xi^*)\end{pmatrix}, \quad (B5)$$

for the state $D(\xi)S(r)|SCS_{\pm}(\alpha_{SCS})\rangle$, where $r$ is a squeezing parameter and $\xi$ is an arbitrary amplitude of displacement. Rewriting the Wigner functions of the even/odd DSSCSs (B5), we have the following

$$W_{\pm DSSCS}(\alpha) = N_{\pm}(\alpha_{SCS}) \left(W_0(\alpha) + W_{-0}(\alpha) \pm 2X_{\alpha_{DSSCS}}(\alpha)\right), \quad (B6)$$

$$W_0(\alpha) = \frac{2}{\pi}\exp\begin{pmatrix}-2\left(\frac{x-x_\alpha}{\exp(r)} - x_{SCS}\right)^2 - \\ 2\left(\frac{p-p_\alpha}{\exp(-r)} - p_{SCS}\right)^2\end{pmatrix}, \quad (B7)$$

$$W_{-0}(\alpha) = \frac{2}{\pi}\exp\begin{pmatrix}-2\left(\frac{x-x_\alpha}{\exp(r)} + x_{SCS}\right)^2 - \\ 2\left(\frac{p-p_\alpha}{\exp(-r)} + p_{SCS}\right)^2\end{pmatrix}, \quad (B8)$$

$$X_{\alpha_{DSSCS}}(\alpha) = \frac{2}{\pi} \exp\begin{pmatrix}-2\frac{(x-x_\alpha)^2}{\exp(2r)} - \\ 2\frac{(p-p_\alpha)^2}{\exp(-2r)}\end{pmatrix}, \quad (B9)$$

$$\cos\left(4\left(p_{SCS}\frac{x-x_\alpha}{\exp(r)} - x_{SCS}\frac{p-p_\alpha}{\exp(-r)}\right)\right)$$

where $\alpha = x + ip$, $\alpha_{SCS} = x_{SCS} + ip_{SCS}$, $\alpha = x_\alpha + ip_\alpha$.

It follows from above analysis, the outcomes of the optical scheme in Fig. 1(b) are either state (6a) or (6b). It is possible to show the Wigner function of displaced two-level superposition ($N=1$, Eq. (6a)), being the simplest approximation of the DSSCSs, is given by

$$W_{\pm 1}(\alpha) = \frac{1}{1+|a_{\pm 1}|^2}\left(W_0(\alpha) + |a_{\pm 1}|^2 W_1(\alpha) + X_{\pm 01}(\alpha)\right), \quad (B10)$$

$$W_0(\alpha) = Y(\alpha), \quad (B11)$$

$$W_1(\alpha) = Y(\alpha)\left(4(x-x_{In})^2 + 4(p-p_{In})^2 - 1\right), \quad (B12)$$

$$X_{\pm 01}(\alpha) = 2Y(\alpha)\begin{pmatrix}a_{\pm 1}^*(x - x_{In} + i(p - p_{In})) + \\ a_{\pm 1}(x - x_{In} - i(p - p_{In}))\end{pmatrix}, \quad (B13)$$

$$Y(\alpha) = \frac{2}{\pi}\exp\begin{pmatrix}-2(x-x_{In})^2 - \\ 2(p-p_{In})^2\end{pmatrix}, \quad (B14)$$

where $\alpha_{In} = x_{In} + ip_{In}$. The Wigner function of three-level superposition ($N=2$, Eq. (6b)) being the next approximation of the SCSs, is given by

$$W_{\pm 2}(\alpha) = \frac{1}{1+|a_{\pm 1}|^2 + |a_{\pm 2}|^2}$$

$$\begin{pmatrix} W_0(\alpha) + |a_{\pm 1}|^2 W_1(\alpha) + \\ |a_{\pm 2}|^2 W_2(\alpha) + X_{\pm 01}(\alpha) + \\ X_{\pm 02}(\alpha) + X_{\pm 12}(\alpha) \end{pmatrix}, \quad (B15)$$

where

$$W_2(\alpha) = Y(\alpha)$$

$$\begin{pmatrix} 1 + 4(2(x-x_\xi)^2 + 2(p-p_\xi)^2 - 1) + \\ 3 + 4(2(x-x_\xi)^2 - 3)(x-x_\xi)^2 + \\ 4(2(p-p_\xi)^2 - 3)(p-p_\xi)^2 + \\ (1 - 4(x-x_\xi)^2)(1 - 4(p-p_\xi)^2) \end{pmatrix}, (B16)$$

$$X_{02}(\alpha) = 2\sqrt{2} Y(\alpha)$$

$$\left( a_{\pm 2} \begin{pmatrix} (x-x_\xi) - \\ i(p-p_\xi) \end{pmatrix}^2 + c.c. \right), \quad (B17)$$

$$X_{12}(\alpha) = 2\sqrt{2} Y(\alpha)$$

$$\begin{pmatrix} a_{\pm 1} a_{\pm 2}^* \begin{pmatrix} ((x-x_\xi) + i(p-p_\xi)) \\ \begin{pmatrix} 2(x-x_\xi)^2 + 2(p-p_\xi)^2 \\ -1 \end{pmatrix} \end{pmatrix} \\ + c.c. \end{pmatrix}, (B18)$$

$$Y(\alpha) = \frac{2}{\pi} \exp\left(-2(x-x_\xi)^2 - 2(p-p_\xi)^2\right), \quad (B19)$$

where $\xi = \alpha_{In} + \alpha_1 = x_\xi + ip_\xi$ and *c.c.* means complex conjugated. Marginal distributions for the momentum and position are given by $\int W(x,p)dx = \langle p|\rho|p \rangle$ and $\int W(x,p)dp = \langle x|\rho|x \rangle$, respectively.

## LIST OF FIGURES

**Figure 1(a,b)**
(a) Schematic optical scheme used for the construction of the even and odd DSSCSs with high fidelity in dependency on an amplitude of seed coherent state $\alpha_{In}$. The optical scheme involves set of alternate photon additions and displacement operators with corresponding amplitudes. One or two single photon additions are used. Use of two single photon additions with one intermediate displacement operator enables to generate DSSCSs of larger amplitude. (b) Application of the scheme to construction of displacing Hadamard gate. The optical scheme starts with either $|0, \alpha_{HG}\rangle$ or $|0, -\alpha_{HG}\rangle$ and displacement operator $D(\beta)$ is used to obtain corresponding amplitude of seed coherent state $\alpha_{In}$. Then, output of the scheme is either even or odd squeezed SCSs, respectively, shifted relative each other by some value (Table 3). Phase shift and absorbing medium at the beginning and squeezing operator at the end of optical scheme (dashed rectangles) can be used to provide construction of regular Hadamard gate.

**Figure 2(a-d)**
Compassion of the output of the optical scheme in Fig. 1(b) with those states that they approximate. Only phase of basis states varies on opposite. All other parameters are left unchangeable. All data are taken from Table 3. (a) Wigner function $W_{+2}$ of the output state when basic state with an amplitude $\alpha_{HG} = i0.474207$ is used. (b) Wigner function of squeezed even SCS $W_{+SSCS}$ with $\alpha_{SCS} = 1.4$ displaced by $\alpha_+ = -i1.32164$. (c) Wigner function $W_{-2}$ of output state if basic state with an amplitude $\alpha_{HG} = -i0.474207$ is used. (d) Wigner function of squeezed odd SCS $W_{-SSCS}$ with $\alpha_{SCS} = 1.4$ displaced by $\alpha_- = -i2.64334$.

**Figure 3(a-d)**
The same as for figure 2. (a) Wigner function $W_{+2}$ of the output state when basic state with an amplitude $\alpha_{HG} = i0.414998$ is used that approximates (b) Wigner function of squeezed even SCS $W_{+SSCS}$ with $\alpha_{SCS} = 1.5$ displaced by $\alpha_+ = i1.22947$. (c) Wigner function $W_{-2}$ of output state if basic state with an amplitude $\alpha_{HG} = -i0.414998$ is used that enables it to approximate (d) Wigner function of squeezed odd SCS $W_{-SSCS}$ with $\alpha_{SCS} = 1.5$.

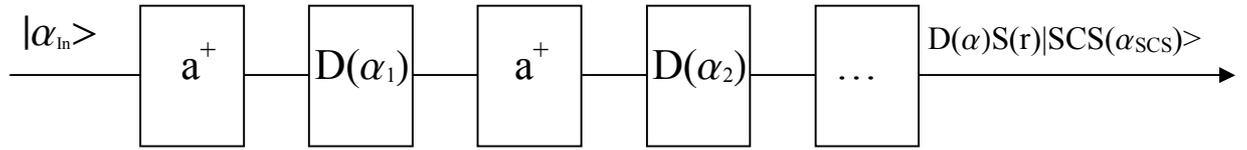

**a)**

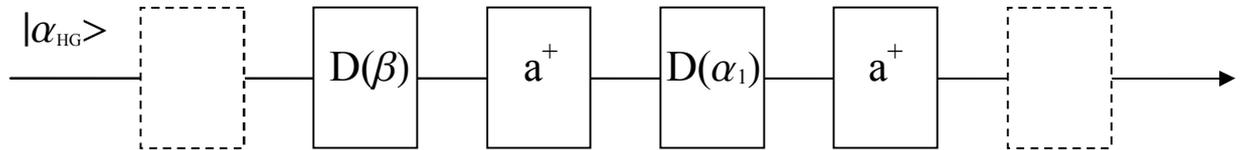

**b)**

**Figure 1(a,b)**

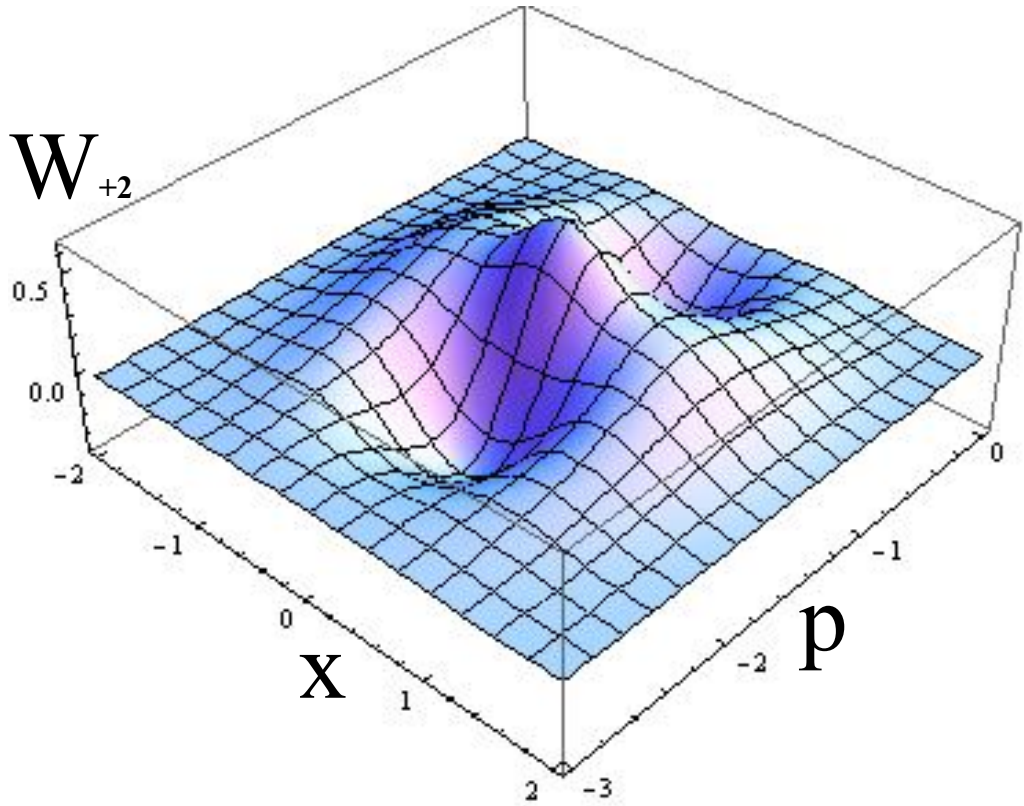

**Figure 2(a)**

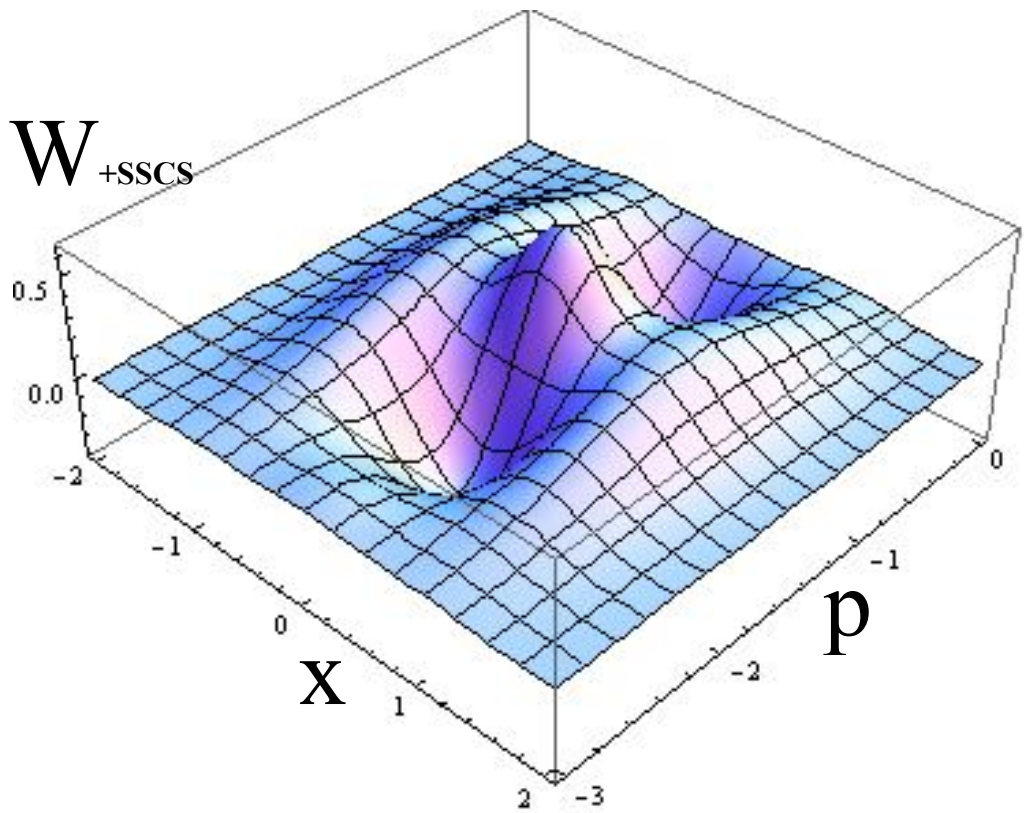

**Figure 2(b)**

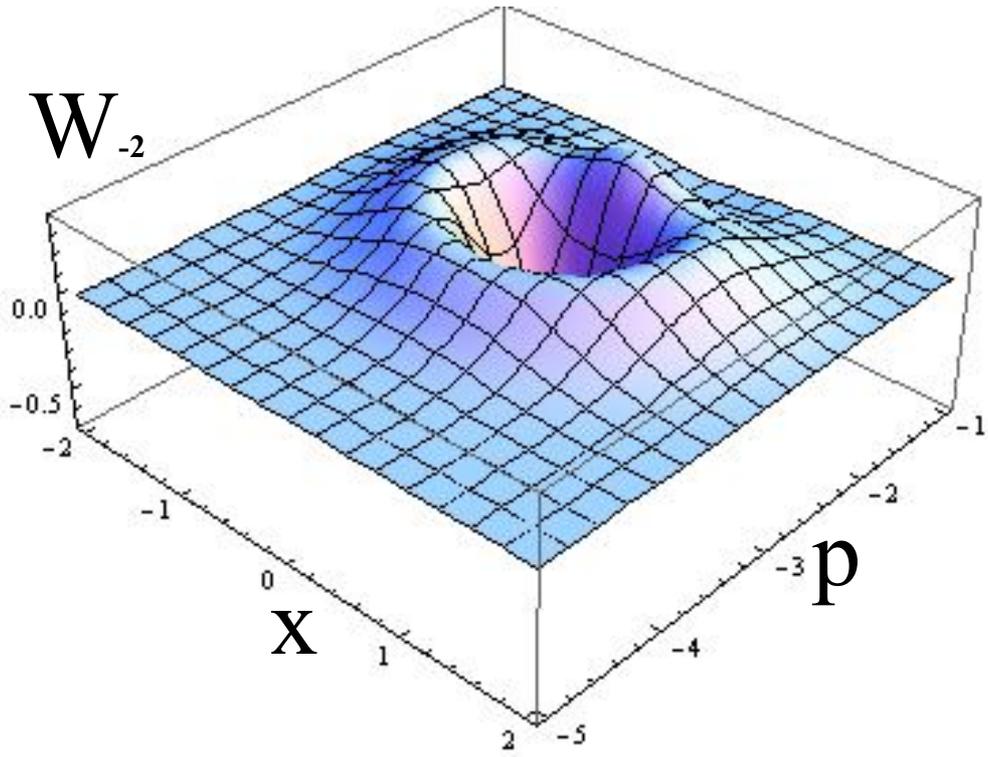

**Figure 2(c)**

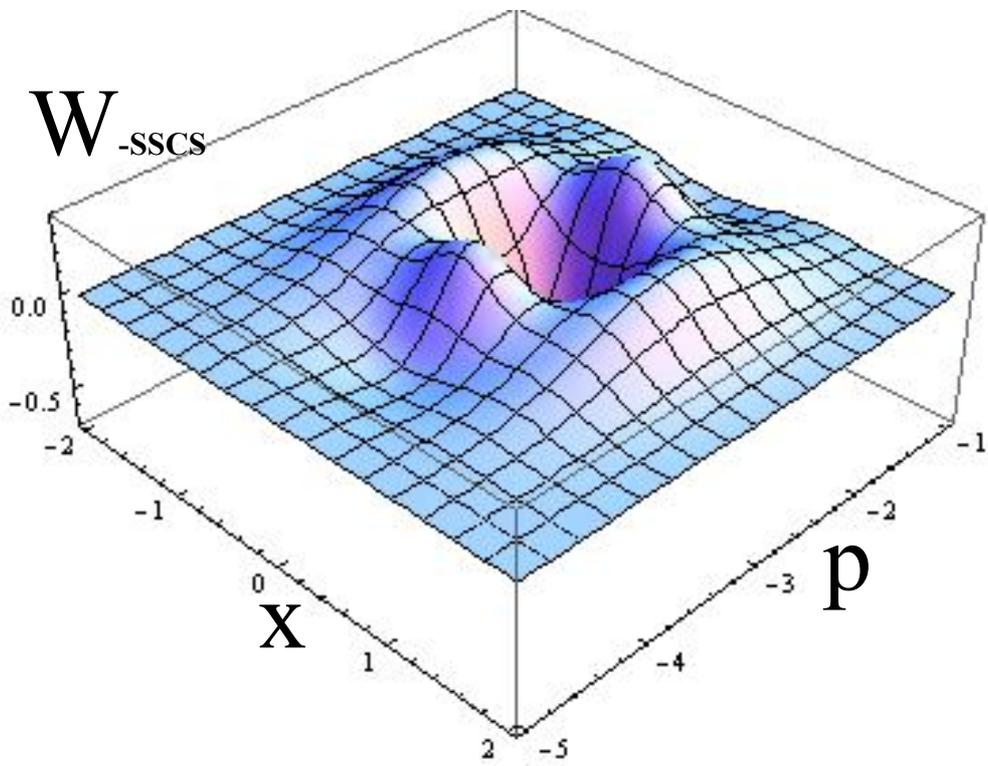

**Figure 2(d)**

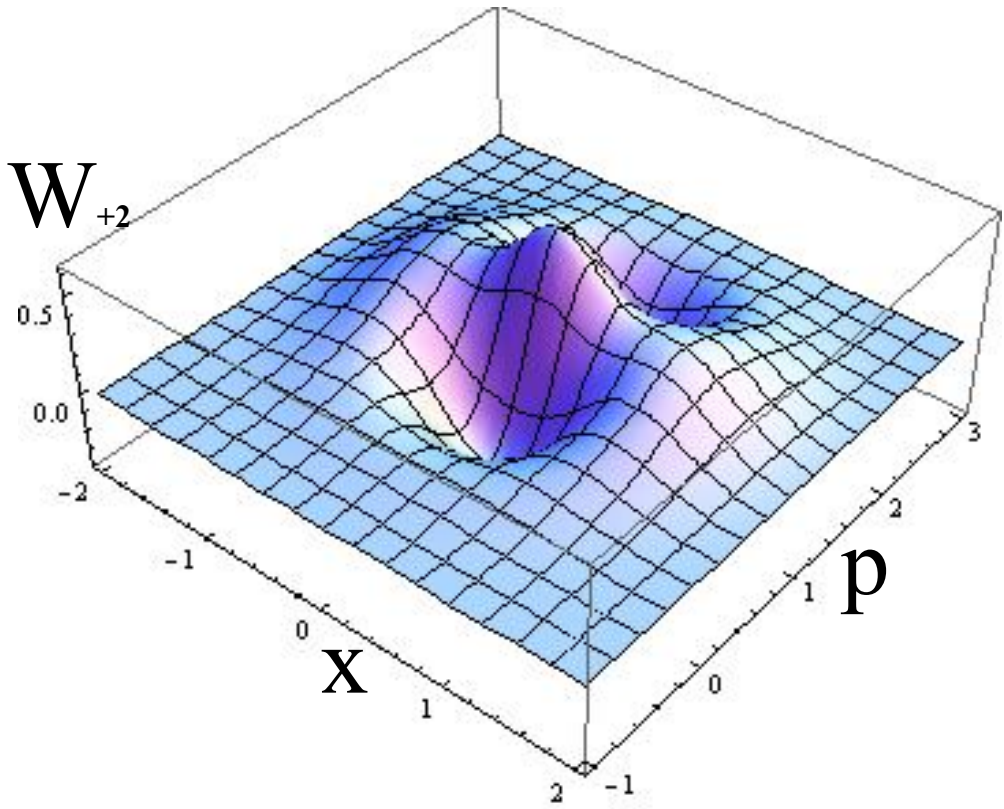

**Figure 3(a)**

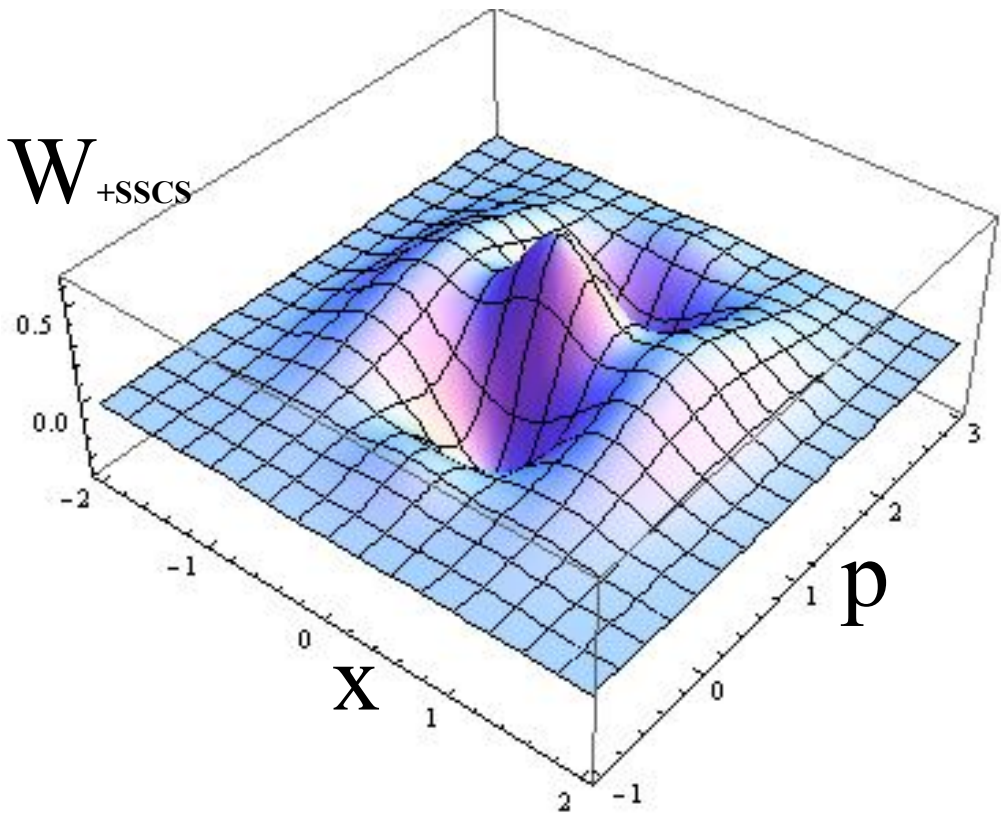

**Figure 3(b)**

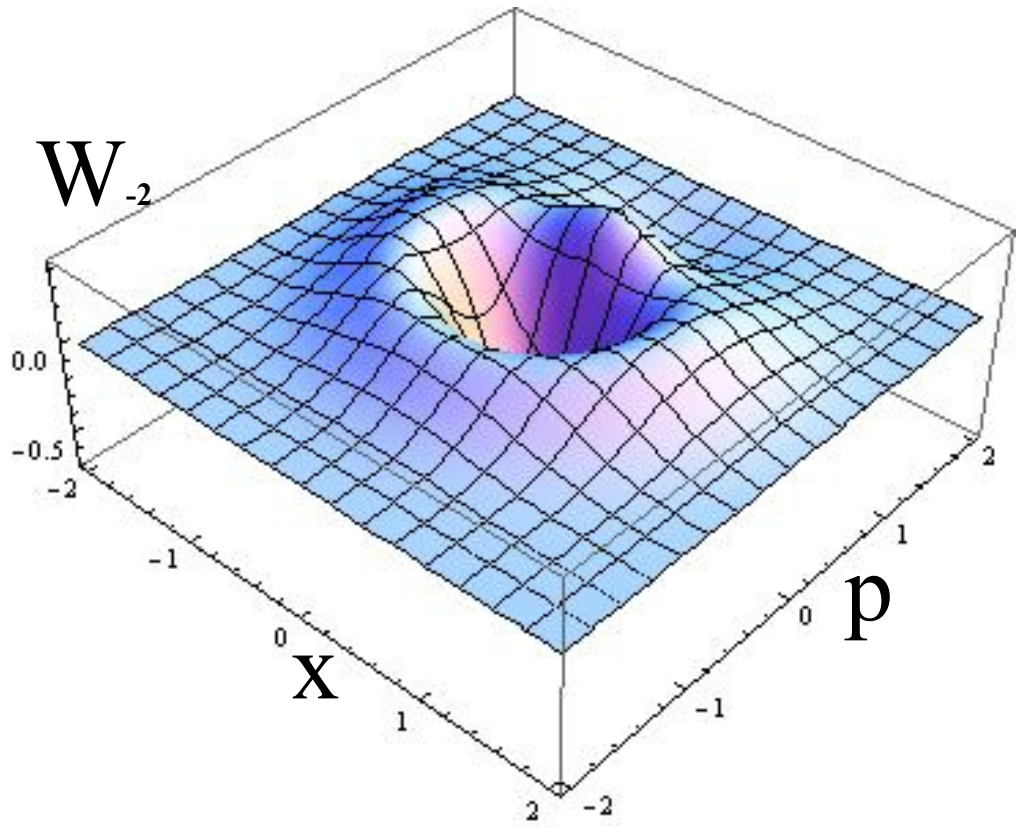

**Figure 3(c)**

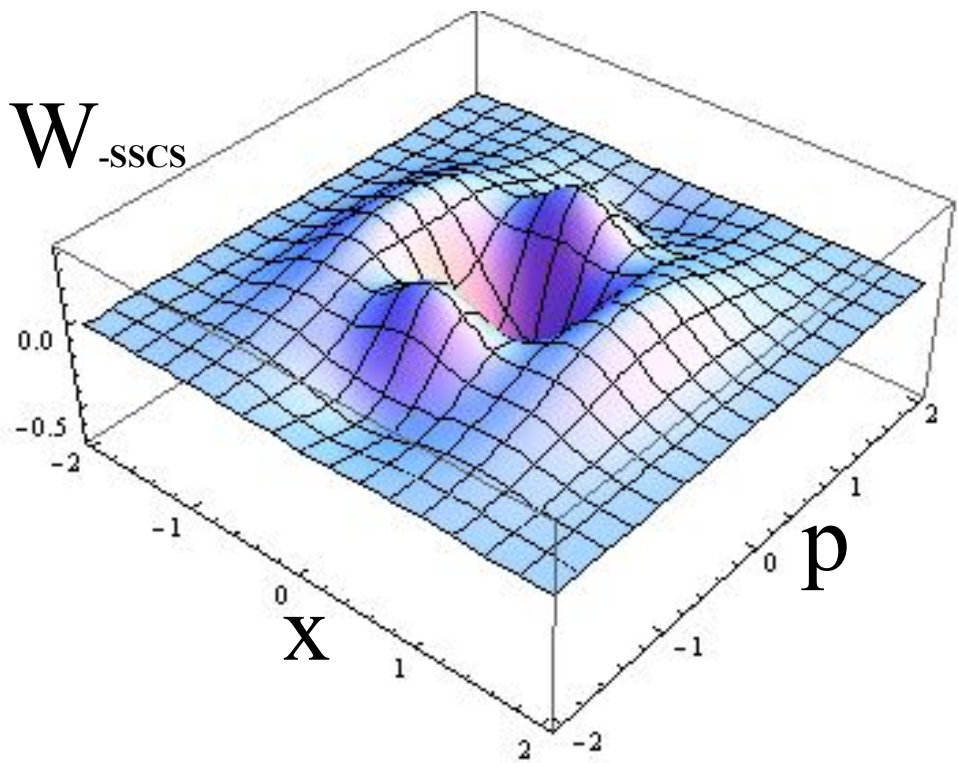

**Figure 3(d)**